\documentclass[a4paper,superscriptaddress,aps,preprintnumbers,amsmath,amssymb,sort&compress,nofootinbib,apj]{revtex4}
\usepackage{graphicx}
\usepackage{amssymb,amsmath,microtype}

\begin{document}
\title{Tibet$^\prime$s window on primordial gravitational waves}

\author{Hong Li}
\affiliation{Key Laboratory of Particle Astrophysics,  Institute of High Energy Physics (IHEP), Chinese Academy of Sciences, 19B Yuquan Road, Shijingshan District, Beijing 100049, China}

\author{Si-Yu Li}
\affiliation{Key Laboratory of Particle Astrophysics,  Institute of High Energy Physics (IHEP), Chinese Academy of Sciences, 19B Yuquan Road, Shijingshan District, Beijing 100049, China}

\author{Yang Liu}
\affiliation{Theoretical Physics Division, Institute of High Energy Physics (IHEP), Chinese Academy of Sciences, 19B Yuquan Road, Shijingshan District, Beijing 100049, China}
\affiliation{University of Chinese Academy of Sciences, Beijing, China}

\author{Yong-Ping Li}
\affiliation{Theoretical Physics Division, Institute of High Energy Physics (IHEP), Chinese Academy of Sciences, 19B Yuquan Road, Shijingshan District, Beijing 100049, China}
\affiliation{University of Chinese Academy of Sciences, Beijing, China}

\author{Xinmin Zhang}
\affiliation{Theoretical Physics Division, Institute of High Energy Physics (IHEP), Chinese Academy of Sciences, 19B Yuquan Road, Shijingshan District, Beijing 100049, China}
\affiliation{University of Chinese Academy of Sciences, Beijing, China}

\maketitle

As an essential part of China's Gravitational Waves Program, the Ali CMB Polarization Telescope (AliCPT) is a ground-based experiment aiming at the Primordial Gravitational Waves (PGWs) by measuring B-mode polarization of Cosmic Microwave Background (CMB). First proposed in 2014 and currently in fast construction phase, AliCPT is China's first CMB project that plans for commissioning in 2019. Led by the Institute of High Energy Physics (IHEP) under the Chinese Academy of Sciences (CAS), the project is a worldwide collaboration of more than fifteen universities and research institutes.

\begin{figure}
\begin{center}
\includegraphics[scale=0.5]{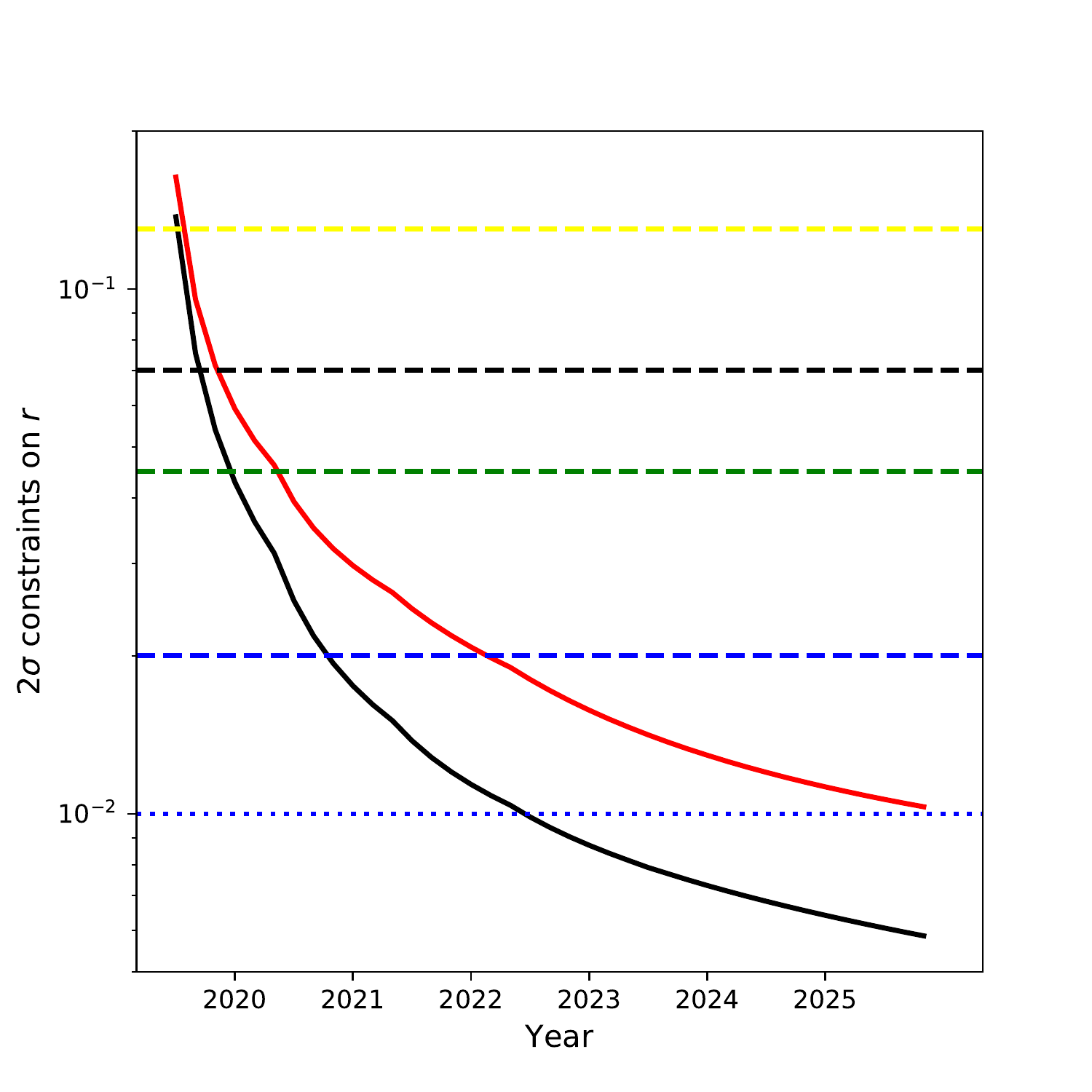}
\caption{Theoretical predictions of three inflation models and the scheduled AliCPT sensitivity of the measurements on $r$. The black and red curves represent the AliCPT $2\sigma$ limits on $r$, where in the simulations we have considered 30\% residual lensing effect, and residual foreground of 1\% (black) and 10\% (red) \cite{hong}. The black dashed line is the current limit from BICEP/Keck Array and Planck collaborations \cite{Ade}.  The yellow dashed line, green dashed line, are the predictions from inflation models with potential function of $\phi^2$ \cite{Linde} and $\phi^{2/3}$ \cite{McAllister}. The blue dashed and dotted lines are for the alpha attractor model \cite{Kallosh} with $\alpha = 7$, $\alpha = 3$ and $n = 1$. In the theoretical calculations, e-folds number is taken to be 60. }
\label{theoretical_predictions}
\end{center}
\end{figure}

Astrophysics and cosmology embrace a new era of observation of gravitational waves. Gravitational waves are emitted by variety of sources. Besides the outburst events from binary mergers detected by LIGO and Virgo collaborations \cite{Abbott}, the PGWs are expected to be generated during the inflationary period of early universe. Inflationary cosmology suggests that the universe has experienced an accelerating phase before the regular thermal expansion. It solves several conceptual issues of the Big Bang cosmology including the flatness, monopole, and horizon problems and also predicts the primordial perturbations originated from the quantum fluctuations of the inflaton field and the space-time. The scalar modes of primordial perturbations can eventually seed the CMB temperature anisotropies and lead to the formation of large-scale structure. And the tensor modes, dubbed as PGWs, will generate the B-mode polarization in the CMB. In Figure \ref{theoretical_predictions}, we plot the theoretical predictions on the tensor-to-scalar ratio $r$ of some representative inflation models and the current experimental limit from the joint analysis by the BICEP/Keck Array and Planck collaborations \cite{Ade}. The parameter $r$ characterizes the amplitude of PGWs. One can see that the measurement of $r$ is powerful in constraining various inflation models, and more precise experiments are required to discriminate them. Moreover, beyond inflation scenario, theorists also proposed different early universe paradigms such as bounce cosmology and cyclic universe. These models generate the PGWs, but give different patterns, which also need CMB polarization experiments like AliCPT for differentiation. 

The major existing ground-based CMB experiments reside in southern-hemisphere, for example, the Atacama Cosmology Telescope (ACT) \cite{Sherwin} and POLARBEAR/Simons Array \cite{Ade2} in Chile, the South Pole Telescope (SPT) \cite{Benson} and BICEP \cite{Ade3} series at the South Pole. An experiment in the northern-hemisphere is essentially needed to achieve the full sky coverage.

\begin{figure}
\begin{center}
\includegraphics[scale=0.55]{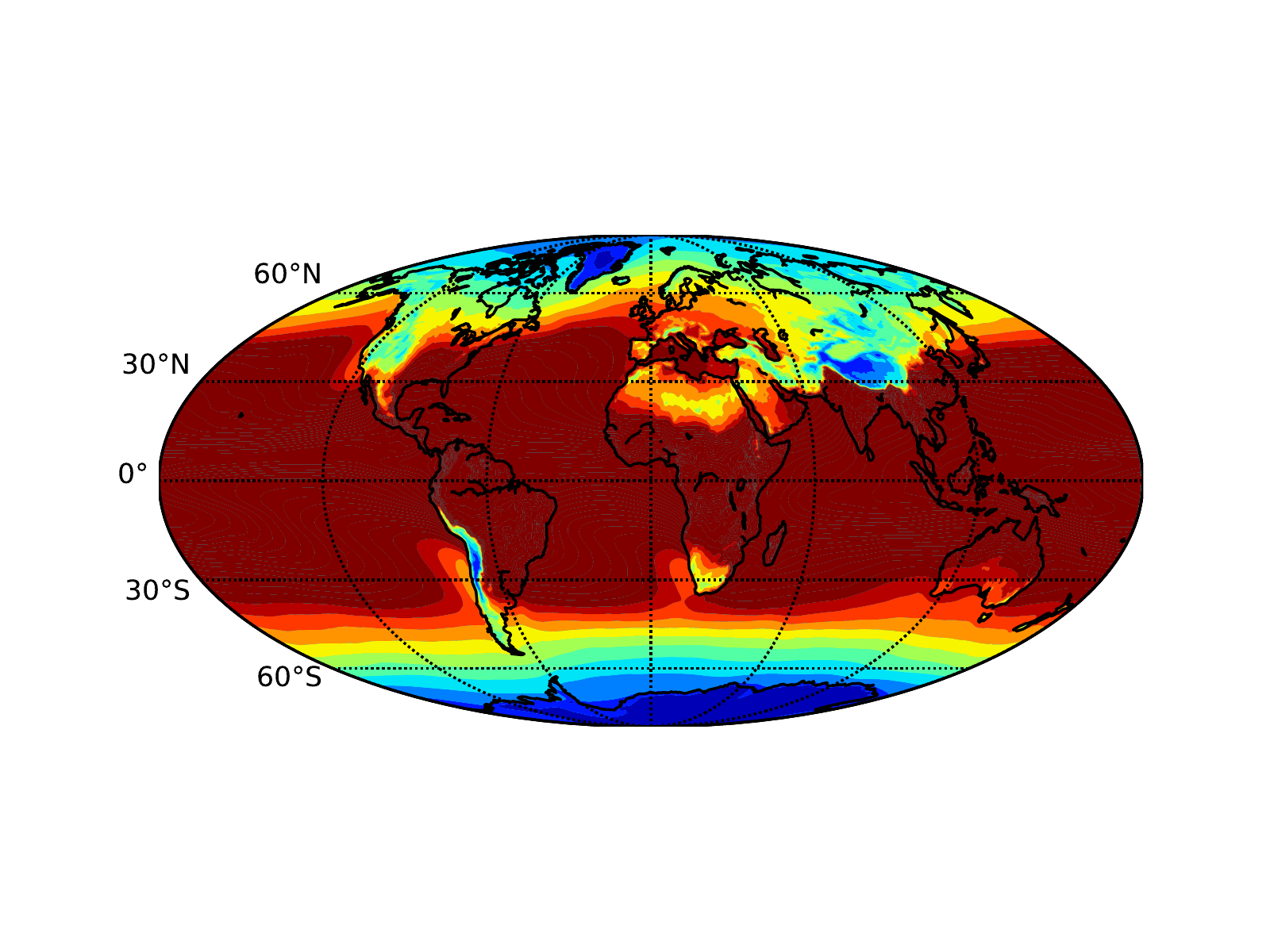}
\includegraphics[scale=0.4]{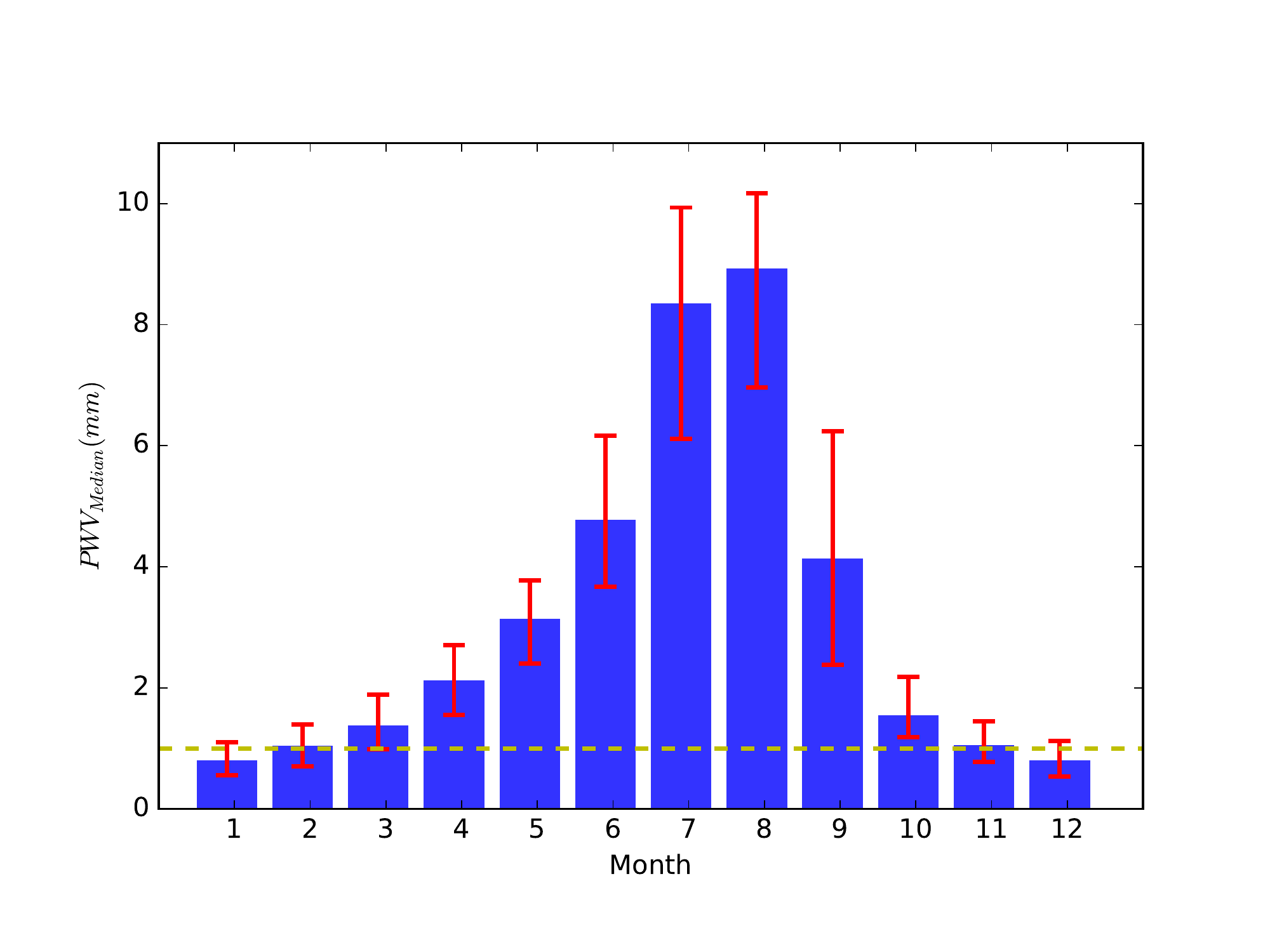}
\caption{Left: Global distribution of mean PWV calculated with MERRA-2 reanalysis data from 2011.7 to 2016.7 \cite{Bosilvoich}, taken from the reference \cite{Ping}. Four regions including Tibetan Plateau in China, Greenland, Atacama Desert in Chile and Antarctica have the lowest water vapor.
Right: Monthly distribution of the median PWV of the site for AliCPT \cite{Ping}. The yellow dashed line represents the PWV value of one millimeter, the error bar means the quartiles of each month. The optimal observing period is from October to March.}
\label{PWV}
\end{center}
\end{figure}

For ground-based CMB experiments, the atmospheric condition is a severe concern. The left panel of Figure \ref{PWV} shows the global distribution of precipitable water vapor (PWV), which indicates that four regions are optimal for ground-based CMB observations. They are the Atacama Desert, Chile and Antarctica in the southern hemisphere, and Tibetan Plateau and Greenland in the north. The AliCPT is located on a hilltop ($32^{\circ}18'38''N, 80^{\circ}1'50''E$) in the Ngari Prefecture, Qinghai-Tibetan Plateau, at an altitude of 5250 meters. The Himalayas is to its southwest and runs from northwest to southeast, separating the Ngari prefecture from the Indian subcontinent as well as the Indian Ocean. As a result, the wet air from the Indian Ocean is largely reduced. Consequently, the air of AliCPT site is thin and dry around winter. In the right panel of Figure \ref{PWV} we present its averaged monthly PWV distribution. As one can see, from October to March, the median PWV is equal to or lower than one millimeter, indicating excellent conditions for CMB observations at frequencies of 95 and 150 GHz.

\begin{figure}
\begin{center}
\includegraphics[scale=0.45]{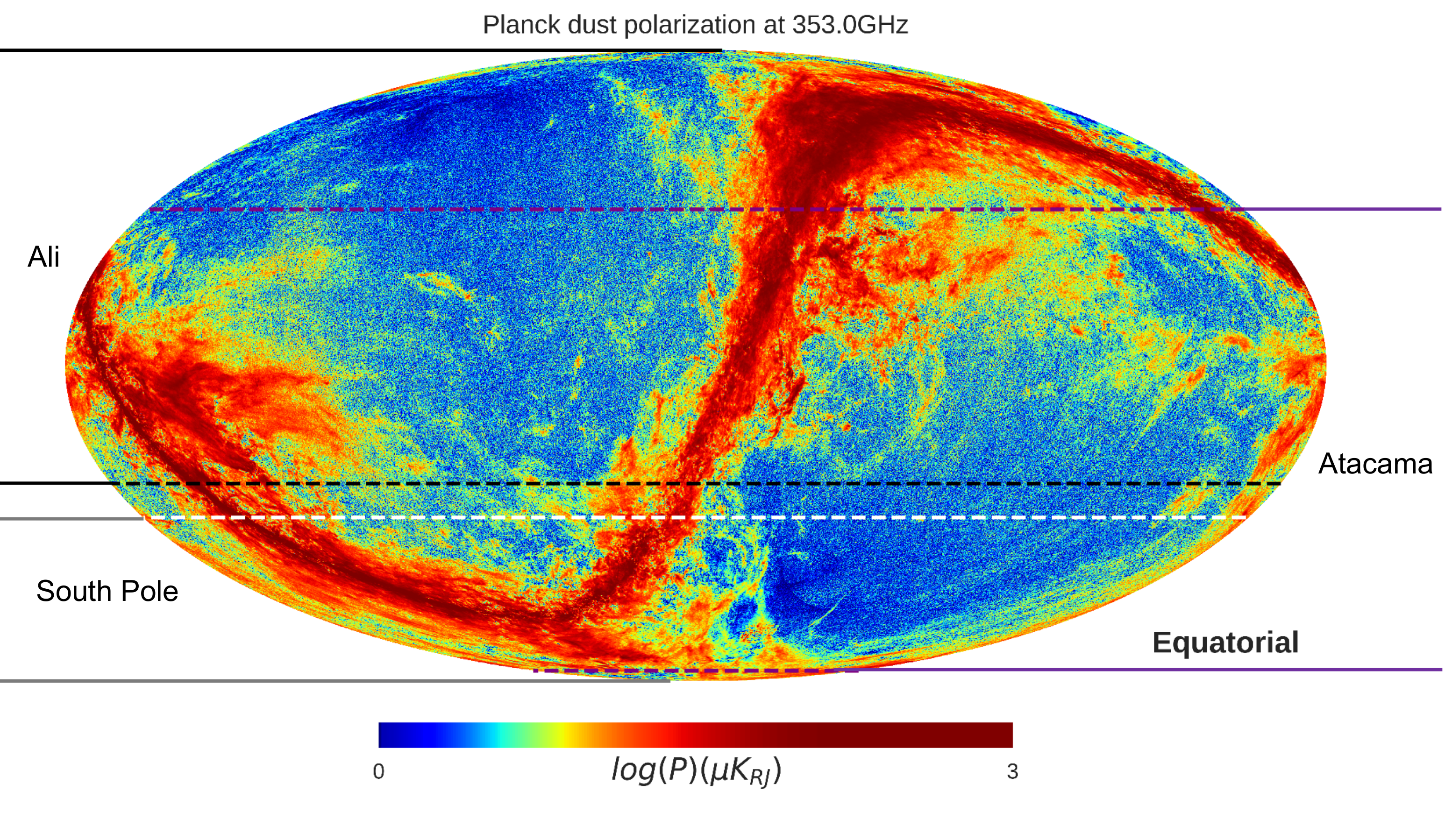}
\caption{Observable sky of AliCPT on the basis of equatorial coordinate. The background is the distribution of polarized dust emission at 353GHz obtained by Planck \cite{Adam}. Observable regions of AliCPT, Atacama and South Pole are shown as the corresponding areas between black, purple and grey lines. The combination of AliCPT with CMB experiments in South Pole and Atacama can realize a full sky coverage.}
\label{observable_sky}
\end{center}
\end{figure}

Compared to the current major ground-based missions, the scanning area of AliCPT is unique due to its location in the northern hemisphere. In Figure \ref{observable_sky} and Table \ref{table:site}, we show the sky coverage of AliCPT and its comparison with those of other CMB-observing ground-based facilities. The AliCPT will complement the experiments in the South Pole and Chile's Atacama, and cover the 'northern galactic hole', an important sky patch with very low foreground contamination.

\begin{table}
\caption {Information on the observational sites of some typical CMB missions worldwide \cite{Ping}\cite{Kuo}. AliCPT(5250) refers to the current AliCPT site with an altitude of 5250 meters. AliCPT(6000) has an altitude of around 6000 meters, which is still within the Ngari prefecture, and has been considered for the site of the forthcoming project. Dome A refers to the site in the Dome A, Antarctica that is possible for the future CMB observation in China’s Kunlun Station.}
\label{table:site}
\begin{center}
   \begin{tabular}{ c  | c | c | c | c | c  }
   \hline			
    Site  & Height(m) & Time range & PWV(mm) & Sky range & Observable fraction (\%) \\
   \hline
 $\text {AliCPT(5250)}$  & 5250 & Oct. - Mar. & 1.07 & Whole North + Part South & 70\\
  \hline
  $\text {AliCPT(6000)}$ & 6000 & Oct. - Mar. & 0.62 & Whole North + Part South & 70\\
  \hline
  South Pole(BICEP3) &2835 & Apr. - Sep. & 0.27 & Part South & 20\\
  \hline
  Atacama(POLARBEAR) & 5200 & Apr. - Sep. & 0.85 & Whole South + Part North & 80\\
  \hline
  Dome A & 4093 & Apr. - Sep. & 0.12 & Part South & 25\\
  \hline

  \end{tabular}
\end{center}
\end{table}

The Ali Observatory and Ngari Gunsa airport are found nearby the AliCPT site, and hence the necessary infrastructure and convenient transportation can be guaranteed. In addition, the observatory site is about 30 kilometers from Shiquanhe town, the regional capital of Ngari prefecture with a population of about twenty thousand, which will provide good living conditions for observers.

The AliCPT project has two planned stages. The first stage will focus on the CMB polarization telescope, dubbed AliCPT-1, which is a dichroic refractor operating at 95 and 150GHz frequency bands with Transition Edge Sensor (TES) bolometers and Superconducting Quantum Interference Devices (SQUIDs) readouts. Sensors and their readouts will be packaged into highly integrated modules, and each contains 1,704 TES sensors. AliCPT-1 will include four modules and its number of detectors will reach 6,816. The second stage, AliCPT-2, will begin in 2020. In this stage, four additional modules will be installed each year until 2022, and the total number of detectors will reach more than 20,000.

The leading scientific goal of AliCPT is to search for the PGWs. In Figure \ref{theoretical_predictions}, we plot the scheduled sensitivity of the measurement on $r$. We can see that with the initial two years' observations until 2022, the constraint on $r$ would be pushed down to $0.01$ at $2\sigma$. In the associated simulations, we have considered the residual foreground and lensing effects \cite{hong}. By 2025, this sensitivity will reach $0.006$, more than one order of magnitude smaller than the current limit $r<0.07$ \cite{Ade}. These high precision measurements will be crucial to test the inflation models as shown in Figure \ref{theoretical_predictions}.

The measurements of CMB B-mode will also be important to the test of the fundamental CPT (Charge conjugation, Parity and Time reversal) symmetry. With a CPT violating Chern-Simons term, the polarized direction of CMB photon will rotate. This rotation will partly convert the CMB E-mode into the B-mode. Current upper bounds on this rotation angle from space and ground-based CMB experiments, including Planck, WMAP, BICEP and POLARBEAR, are around $1^{\circ}$ \cite{hong}. AliCPT is expected to improve this limit by two orders of magnitude down to ${0.01}^{\circ}$ with 3 years of observations \cite{hong}. Probing such a violation of the CPT symmetry is significant in exploring new physics beyond the standard models of cosmology and particle physics.

In addition to the sciences discussed above, AliCPT will also study the galactic foreground, the hemispherical polarization asymmetry and the correlation between CMB polarization and large-scale structures.

The AliCPT project is funded by National Natural Science Foundation of China, Ministry of Science and Technology of China, and Chinese Academy of Sciences. The site construction started in the early 2017 and the telescope is scheduled to begin the survey in 2020.


\begin{thebibliography}{999}
\bibitem{Abbott}
Abbott, B. P. et al. 
Phys. Rev. Lett. 116 061102 (2016)


\bibitem{hong}
Li, H., et al.
arXiv:1710.03047 (2017)

\bibitem{Ade}
Ade, P. A. R., et al. 
Phys. Rev. Lett., 116, 031302 (2016)

\bibitem{Linde}
Linde, A. D. 
Phys. Lett. B 129, 177 (1983)

\bibitem{McAllister}
McAllister, L.,et al. 
JHEP 1409, 123 (2014)

\bibitem{Kallosh}
Kallosh, R. et al. 
JHEP 1311, 198 (2013)


\bibitem{Sherwin}
Sherwin, B. D., et al. 
Phys. Rev. D 95, 123529 (2017) 

\bibitem{Ade2}
Ade, P. A. R., et al., 
Astrophysical Journal, 794, 171 (2014)

\bibitem{Benson}
Benson, B. A., et al. 
Proceedings of SPIE, 9153 (2014)

\bibitem{Ade3}
Ade, P. A. R., et al., 
Phys. Rev. Lett. 112, 241101 (2014) 

\bibitem{Bosilvoich}
Bosilvoich, M., G. et al. 
NASA Technical Report Series on Global Modeling and Data Assimilation, 43, 20

\bibitem{Ping}
Li, Y. P. et al. 
arXiv:1709.09053 (2017)


\bibitem{Adam}
Adam, R. et al. 
Astron. Astrophys. 594, A10 (2016)

\bibitem{Kuo}
Kuo, C. L., 
arXiv:1707.08400 (2017)






\end{thebibliography}
\end{document}